\documentclass[PAPER, cernpreprint=true, atlasdraft=false, USenglish, texmf, orcidlogo]{atlasdoc}

\usepackage[subcaption]{atlaspackage}
\usepackage{braket}
\usepackage{makecell}
\usepackage{atlasbiblatex}
\usepackage{atlasphysics}
\graphicspath{{logos/}{figures/}}
\usepackage{ANA-HIGP-2024-25-PAPER-defs}

\AtlasTitle{
Measurements of $Z$-boson pair entanglement in decays of Higgs bosons at the ATLAS experiment
}
\AtlasAbstract{Entanglement is a key property of quantum systems. In this Letter the first measurements of quantum entanglement between spins in pairs of $Z$ bosons are reported, using proton--proton collision data from the Large Hadron Collider (LHC) at center-of-mass energies of 13\,TeV and 13.6\,TeV,  recorded with the ATLAS detector. Measurements of angular observables sensitive to $ZZ^*$ spin-density-matrix elements in the $H\rightarrow ZZ^* \rightarrow \ell^+\ell^-\ell^+\ell^-$ process yield coefficients $C_{2,1,2,-1} = -0.71 \pm 0.45$ and $C_{2,2,2,-2}=0.08 \pm 0.44$, consistent with their Standard Model predictions. A complementary hypothesis test using the full angular  distribution, and relying on several Standard Model assumptions in the decays,  provides substantially higher sensitivity to quantum correlations and disfavors the separable-state hypothesis at a significance of 4.7~standard deviations (expected $4.9\sigma$) relative to the entangled Standard Model hypothesis. These results provide strong evidence of quantum entanglement between massive bosons (spin qutrits) at the electroweak scale.
}

\AtlasRefCode{HIGP-2024-25}

\PreprintIdNumber{CERN-EP-2026-061}

\AtlasJournalRef{\PRL 137 (2026) 111804}
\AtlasDOI{10.1103/y1nh-1b82}

\AtlasCoverEgroupAnalysisTeam{atlas-higp-2024-25-analysis-team@cern.ch}

\hypersetup{pdftitle={ATLAS document},pdfauthor={The ATLAS Collaboration}}

\begin{document}
\begin{refsection}

\maketitle

Quantum entanglement is one of the fundamental features of quantum mechanics~\cite{Einstein:1935rr,Schrodinger1935,Bell:1964kc},
lying at the heart of quantum information science and quantum field theory~\cite{Bennett2000,Nielsen2000,Marciniak2022,Horodecki:2009zz,Casini:2022rlv}.
In quantum mechanics, an entangled system cannot be described as a product of its subsystems.\footnote{More precisely the condition for a density matrix $\rho$ to be entangled is that it is not possible to express it  as a convex sum of tensor products $\rho=\sum_i p_i\,\rho_A\otimes\rho_B$ of the density matrices $\rho_A$ and $\rho_B$ of the subsystems $A$ and $B$ for $p_i\geq0$ and $\sum_ip_i=1$.}
Its joint state exhibits correlations that cannot be reproduced by any non-entangled quantum model.

In recent years, collider-based studies have been proposed which would probe entanglement at the highest experimentally accessible energy scales~\cite{Afik:2020onf,Afik:2022kwm,Aguilar-Saavedra:2022wam,Ashby-Pickering:2022umy,Fabbrichesi:2023cev},
building on techniques first developed for optical and atomic systems~\cite{Aspect1982,Hagley1997,Steffen2006,Pfaff2013}.
Entanglement has also been observed in flavor in systems of $K$ mesons~\cite{KLOE-2:2021ila} and $B$ mesons~\cite{Belle:2007ocp}.
Analyses at CERN's Large Hadron Collider (LHC) have demonstrated the feasibility of tests of spin entanglement in top-quark pair production, culminating in the recent observation by the ATLAS and CMS collaborations~\cite{TOPQ-2021-24,CMS-TOP-23-001,CMS-TOP-23-007}.
As well as being intrinsically of interest, such measurements offer sensitivity to possible new particles and fields, with quantum tomography of the polarization density matrix providing a systematic basis for constraining anomalous couplings~\cite{Fabbrichesi:2023jep,Aoude:2023hxv,Maltoni:2024tul} and, more broadly, for probing deviations from Standard Model dynamics~\cite{Altomonte:2024upf}.
Recent reviews of the broader collider programme in quantum information measurements can be found in Refs.~\cite{Barr:2024djo,Afik:2025ejh}.
Extending such measurements to systems with bosonic particles enables tests of quantum correlations among massive vector fields~\cite{Aguilar-Saavedra:2022wam,Ashby-Pickering:2022umy,Fabbrichesi:2023cev}, and provides a probe of the electroweak symmetry-breaking sector responsible for generating the masses of elementary particles~\cite{Higgs1964a, Higgs1964b, EnglertBrout1964, GuralnikHagenKibble1964, Kibble1967}.

The decay of the Higgs boson into a pair of $Z$ bosons provides an ideal laboratory for such studies.
In the Standard Model, the Higgs boson is a scalar particle %
and is even under the charge‑conjugation (C) and the parity (P) transformations ($J^{PC} = 0^{++}$).
The $Z$ is a massive spin-1 boson with three possible spin states, called spin qutrit, and so each $Z$ spin state can be labeled $\ket{+}$, $\ket{0}$,  or $\ket{-}$ according to its helicity.
Since the mass of the Higgs boson is smaller than twice that of the $Z$ boson,
the $H\rightarrow ZZ$ decay involves at least one off-shell (virtual) $Z$ boson denoted $Z^*$.
Virtual $Z^*$ bosons decaying into highly relativistic leptons are also spin qutrits, spanning the
same triplet of helicity states.
The Higgs boson decay into four charged leptons via two spin-1 bosons, $H\rightarrow ZZ^* \rightarrow \ell^+\ell^-\ell^+\ell^-$,
therefore produces a system that can be described by three dominant $Z$-pair helicity amplitudes:
$\ket{+-}$, $\ket{00}$, and $\ket{-+}$.
The corresponding two-boson quantum spin state,
$\ket{ Z Z} = a\ket{+-} + b\ket{00} + c\ket{-+}$,
is expected to be non-separable due to the coherent superposition of helicity components.
The presence of off-diagonal elements in the associated 9$\times$9 spin-density matrix (SDM) therefore constitutes
a direct signature of entanglement between the spins of the two $Z$ bosons~\cite{Aguilar-Saavedra:2022wam}.

This study presents the first measurement of quantum entanglement between massive vector bosons at the electroweak scale.
Distinctive features make such measurements at high-energy colliders fundamentally different
from those in atomic or photonic systems.
First, the LHC probes energies at the electroweak scale,
allowing tests of entanglement for short-lived, massive particles.
Second, in the context of the Standard Model of particle physics, collider experiments can reconstruct the spins of virtual, off-shell bosons through
angular correlations of their decay leptons because
the chiral nature of weak decay allows to relate these angular distributions to the spin state of the parent particles.
A further advantage of the Higgs decay process studied in this Letter is that it offers a clean, fully reconstructible final state with minimal background.

This Letter presents the analysis of Higgs-boson decays into four charged leptons recorded with the ATLAS detector~\cite{PERF-2007-01} at the LHC.
It reports measurements of entanglement-sensitive angular distributions of $ZZ^*$ decay products and the corresponding average values that are related to two entanglement-sensitive SDM coefficients $C_{2,1,2,-1}$ and $C_{2,2,2,-2}$~\cite{Aguilar-Saavedra:2022wam}.
The measured average values are interpreted within the Peres-Horodecki criterion~\cite{Peres:1996dw,Horodecki:1997vt} for separability.
The measured angular distributions are also employed to test the hypothesis of an entangled $ZZ^*$ system as predicted by the Standard Model against a null hypothesis of a non-entangled system in which both $Z$ bosons are longitudinally polarized ($\ket{00}$).
For a CP‑even spin‑0 Higgs boson, as predicted by the Standard Model, this purely longitudinally polarized system is the only configuration that can yield a non‑entangled state~\cite{Aguilar-Saavedra:2022mpg}.
Both approaches allow for a direct experimental test of quantum entanglement of spins in a fundamental bosonic interaction.
The first approach yields a direct measure of entanglement via the spin‑density matrix elements; however the spread of the angular distribution diminishes the statistical accuracy that can be achieved using existing event samples. The second approach achieves greater sensitivity by using the full distribution, with some increased model dependence, as it relies on a comparison between two specific hypotheses.

The analysis rests on several assumptions,\label{sec:assumptions} each supported by both Standard Model expectations and prior experimental results.
First, the pairs of $Z$ bosons produced in Higgs-boson decays are assumed to form a scalar state, consistent with existing measurements of the Higgs boson spin and CP~\cite{HIGG-2013-01,CMS-HIG-14-018}.
Second, the angular distributions of the leptons arising from each helicity state of the $Z^{(*)}$ bosons are assumed to follow the Standard Model predictions, enabling the measurement of its spin from those distributions.
Approximate conservation of CP excludes individual transverse $Z$-boson polarizations.
Higher-order corrections in electroweak perturbation theory, including those from processes without $ZZ^*$ intermediate states, are predicted to
affect leptonic angular distributions and can lead to large biases in measurements of some elements of the SDM~\cite{Goncalves:2025mvl,DelGratta:2025xjp}. Dedicated observables are therefore selected that probe particular entanglement-sensitive elements of the matrix that have small ($<$2\%) higher-order corrections~\cite{Aguilar-Saavedra:2025byk}, these corrections being treated as systematic uncertainties.
Finally, the angular distributions of the four-lepton background processes are assumed to be well-described by the Standard Model, an assumption that is then verified with dedicated signal-free control data.
Together, these assumptions\footnote{The interpretation of collider data in non-Quantum Mechanical theories via Bell’s inequality has been discussed in Ref.~\cite{Abel:1992kz}.
The results presented in this paper are not concerned with probing alternative theories to quantum mechanics via Bell inequalities, but rather in establishing the quantum non-separability of the $Z$ bosons spin states in $H \rightarrow ZZ^*$ decays.}
enable the interpretation of the observed angular correlations in $H\rightarrow ZZ^* \rightarrow 4\ell$ decays in terms of quantum spin entanglement.

The analysis uses proton--proton collision data collected with the ATLAS detector~\cite{PERF-2007-01} at the LHC, employing the full Run~2 data set (2015--2018) at a center-of-mass energy of 13~\TeV and partial Run~3 (2022--2024) data at 13.6~\TeV, corresponding to integrated  luminosities of 140~fb$^{-1}$ and 164~fb$^{-1}$, respectively.
ATLAS is a multipurpose cylindrical particle detector composed of several subdetectors with a forward–backward symmetric cylindrical geometry and a solid-angle coverage of almost $4\pi$. It is used to record particles produced in LHC collisions through a combination of particle position  and energy measurements. It consists of  an inner tracking detector (ID) surrounded by a thin superconducting  solenoid providing a 2\,T axial magnetic field, electromagnetic and  hadronic calorimeters, and a muon spectrometer (MS). The MS surrounds the calorimeters and is based on three large superconducting air-core toroidal magnets with eight coils each providing a field integral of between 2.0\,Tm and 6.0\,Tm across the detector.

Signal and background processes are modeled using Monte Carlo event generators with detailed simulation of the detector response based on \textsc{Geant4}~\cite{Agostinelli:2002hh}.
All samples are normalized to the most precise theoretical cross‑sections currently available. The simulated events are weighted to correct for differences between simulated and measured lepton trigger, reconstruction, identification and isolation performance. Background simulations are validated in dedicated signal-free control regions.

\begin{figure}\centering
\includegraphics[width=0.4\textwidth]{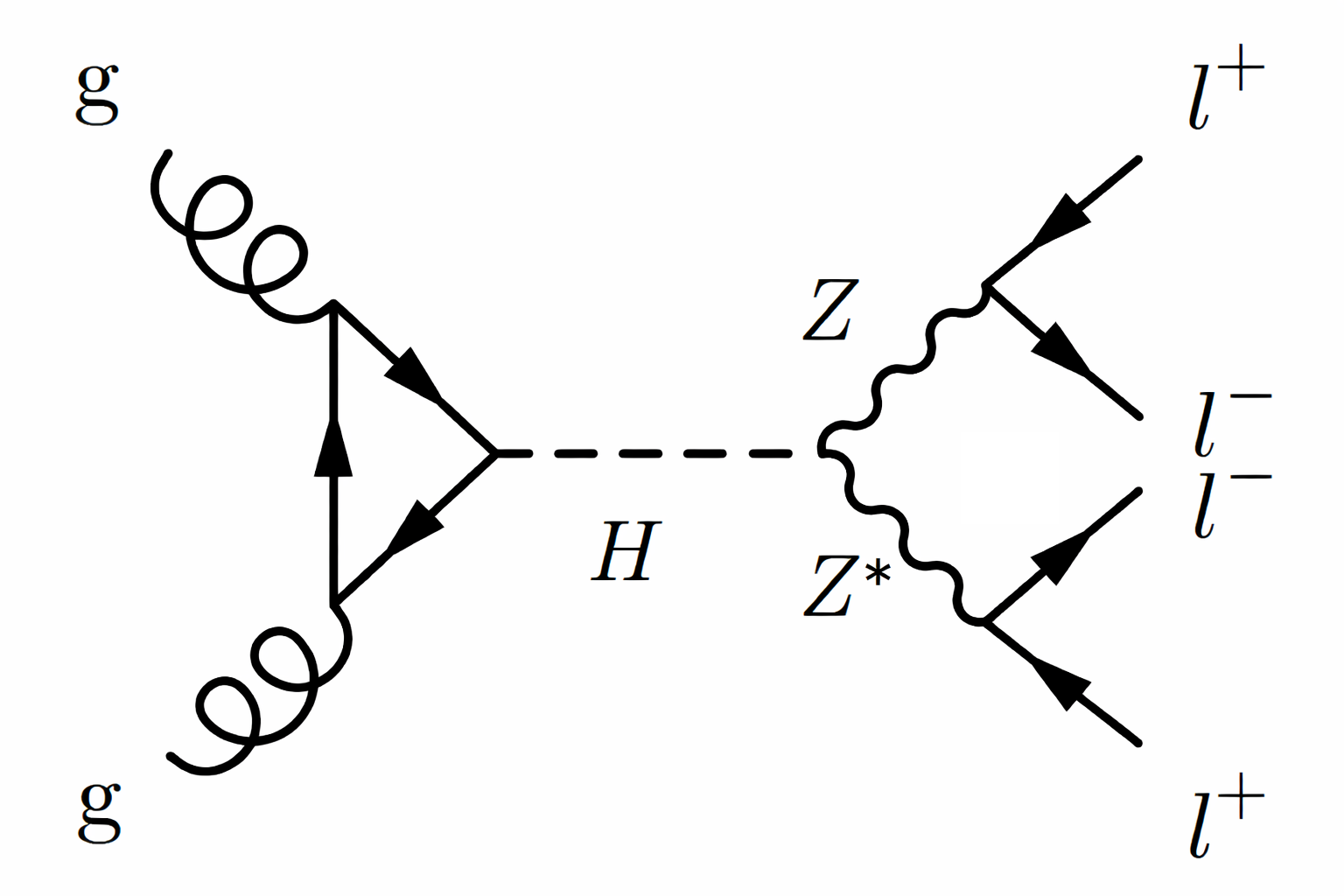}
\caption{\label{fig:Feynman-ggf}
Example of a leading-order Feynman diagram of  the dominant Higgs-boson production process via
gluon–gluon fusion
through a fermion loop and its decay process into two leptonically decaying $Z$ bosons,
$pp\rightarrow H \rightarrow ZZ^* \rightarrow \ell^+\ell^-\ell^+\ell^-$. The superscript (*) refers to a particle that
is off its mass shell.}
\end{figure}

The simulated signal corresponds to Standard Model Higgs boson production via the dominant gluon--gluon fusion process (Figure~\ref{fig:Feynman-ggf}),
vector-boson fusion, and production in association with a $W$ or a $Z$ boson or with one or more third-generation quarks, followed by decay into four charged leptons. This signal is hereafter denoted `QE $H\rightarrow ZZ^*$'.
A null hypothesis corresponding to a non-entangled (separable) quantum system with both $Z$ bosons longitudinally polarized ($\ket{00}$) is simulated and denoted `Non-QE~$H\rightarrow ZZ^*$'.
Backgrounds are dominated by non-resonant diboson production (denoted $ZZ^{(*)}$), with smaller contributions from top quarks (or anti-quarks) produced in association with $Z$ bosons ($ttV+tVV$) and tri-boson production ($VVV$) where $V$ is a $W^\pm$ or $Z$ boson. A subdominant `reducible' background arises from $Z$+jets events and $t\bar{t}$ when events with fewer than four leptons from the interaction point enter the analysis.  Details of the signal and background event simulation
may be found in End Matter~\ref{sec:app:MC}.

The analysis is based on the correlations in angular distributions of charged leptons originating from pairs of $Z^{(*)}$ bosons produced in the decays of Higgs bosons.
Selection criteria are employed following an established ATLAS $H \to ZZ^* \to 4\ell$ event selection strategy~\cite{HIGG-2020-07}.
Events are selected using single- and dilepton triggers~\cite{TRIG-2018-05,TRIG-2018-01} for electrons and muons, which together have an efficiency of about $98\%$ for selecting signal events in the $H\to ZZ^*\to 4\ell$ final state.
Events are required to contain at least four isolated charged leptons ($e$ or $\mu$).
Electrons are reconstructed from electromagnetic clusters matched to particle tracks within the
ID~\cite{EGAM-2021-02} and must have transverse momentum $\pT > 7~\GeV$ and pseudorapidity $|\eta| < 2.47$ (with the region $1.37 < |\eta| < 1.52$ excluded).
Muon candidates are reconstructed using tracks in the MS which within $|\eta| < 2.5$ are matched to ID tracks, and with additional recovery in $|\eta| < 0.1$ using ID tracks matched to MS segments or energy deposits in the calorimeters compatible with a minimum-ionizing particle~\cite{MUON-2018-03}.
Muons are required to have transverse momentum $p_{\mathrm{T}} > 5$~\GeV and $|\eta| < 2.7~(2.5)$ for Run~2~(Run~3) data. Isolation criteria are imposed on the leptons and they are required to originate from a common vertex point.
The three highest-$\pT$ leptons must have transverse momenta exceeding 20, 15, and 10~\GeV, respectively.

\begin{figure}
\centering
\includegraphics[width=0.5\linewidth]{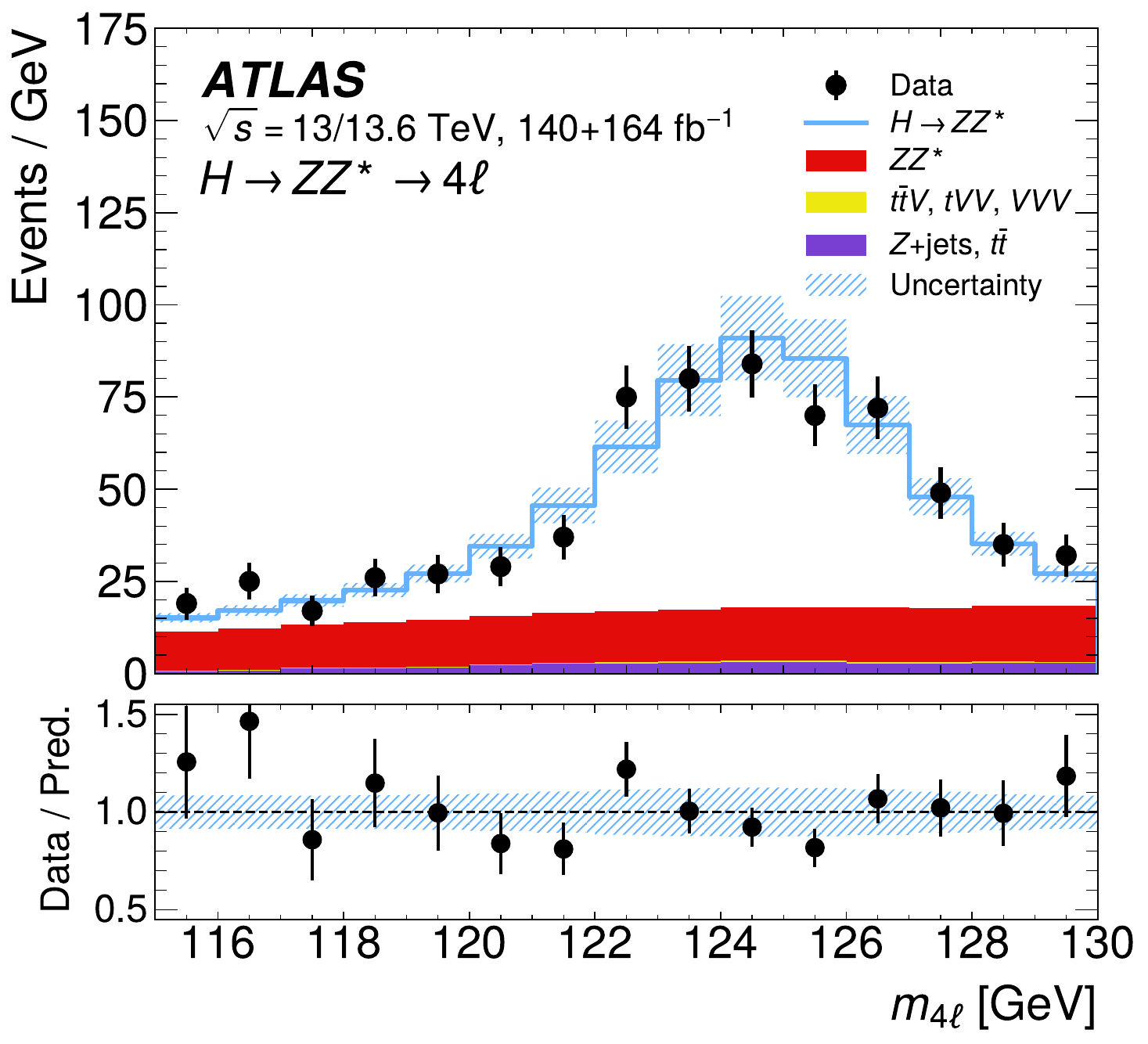}
\caption{\label{fig:m4l}
The observed (full circles) and the expected (histograms) four-lepton invariant mass distribution around the observed Higgs boson resonance. The expectation shows the contributions from $H\rightarrow ZZ^* \rightarrow 4\ell$ and from various background processes as described in the text. The uncertainty bar on data points are Poisson errors, and the uncertainty band includes statistical and systematic contributions added in quadrature.
}
\end{figure}

The analysis is performed for the $4e$, $4\mu$, and $2e2\mu$ channels.
The candidate Higgs boson in each event is reconstructed from two pairs of same-flavor opposite-charge leptons ($e^+e^-$ or $\mu^+\mu^-$). From all valid pairings of the candidate leptons,
the lepton pair with the invariant mass closest to the nominal value of the $Z$ boson mass~\cite{ParticleDataGroup:2024cfk} is denoted $Z_1$, while the second closest is denoted $Z_2$. Any additional leptons  not entering those pairings play no role in the analysis. The invariant mass of these two candidates must satisfy 50\,\GeV{} $< m_{Z_1} <$ 106\,\GeV{} and  12~\GeV{} $ < m_{Z_2} < 115$\,\GeV{}, and the invariant mass of four leptons must be close to the Higgs boson mass~\cite{HIGG-2022-20,ParticleDataGroup:2024cfk}: 115\,\GeV{} $<m_{4\ell} <$ 130\,\GeV.
The observed and expected $m_{4\ell}$ distributions from combined Run~2 and Run~3 data samples are shown in Figure~\ref{fig:m4l}.

\begin{figure}
\centering
\includegraphics[width=0.7\linewidth, trim={0 50 0 0},clip]{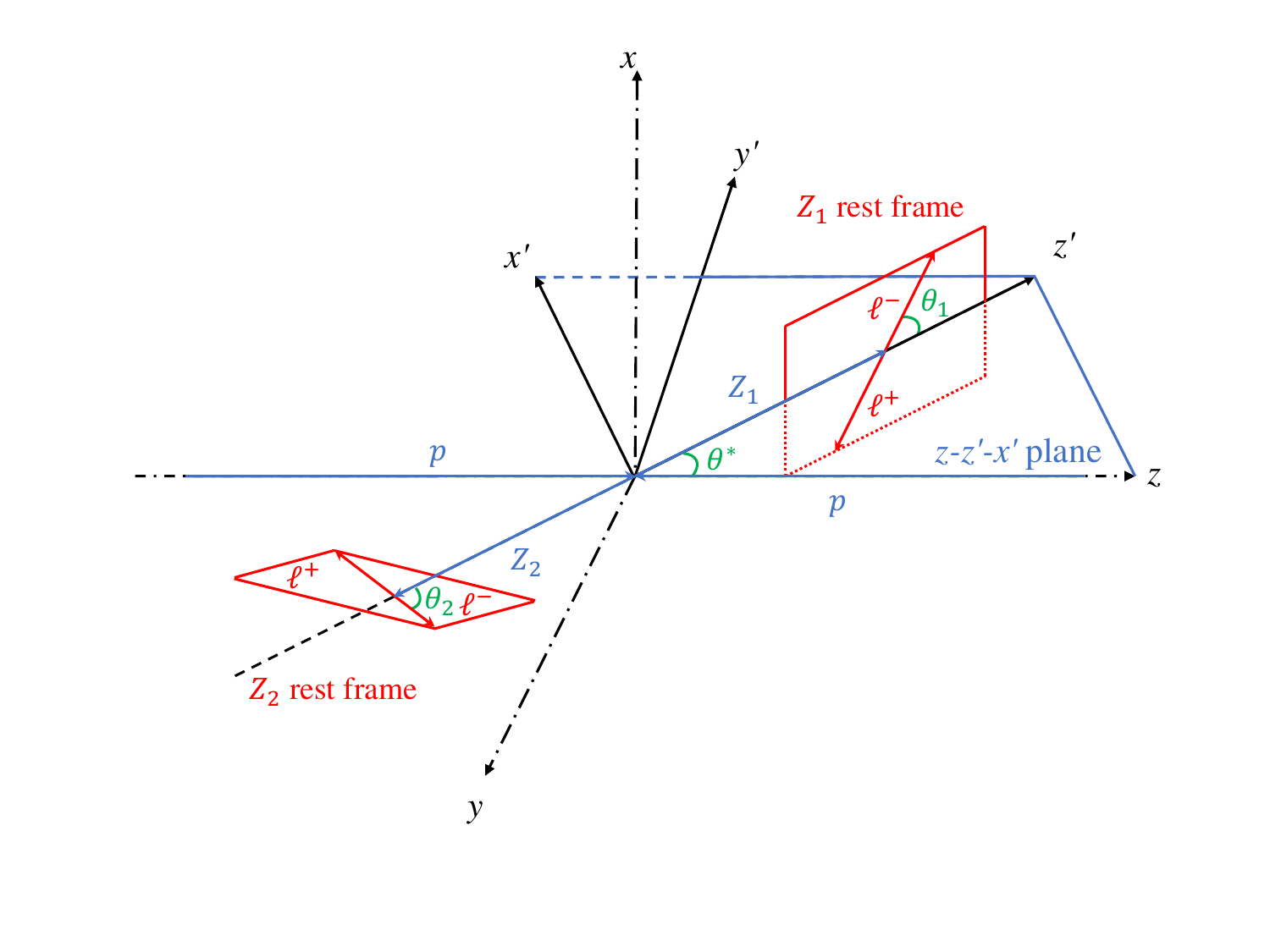}
\caption{\label{fig:geometry} %
An illustration of the decay angle definitions in the $pp\rightarrow H\rightarrow ZZ^* \rightarrow 4\ell$ process. The azimuthal angles $\phi_1$ and $\phi_2$ are not shown in the figure and are defined in the conventional manner for right-handed coordinate systems. The unprimed frame is the laboratory frame with the $z$-axis along the beam direction. The primed frame is that in which the polarization of $Z$ bosons is defined.
}
\end{figure}

For each selected event, the lepton emission directions are defined relative to separate coordinate systems constructed in the rest frame of each parent $Z$ boson~\cite{HIGG-2013-17} as illustrated in Figure~\ref{fig:geometry}.
The $z$-axis in both is aligned with the momentum direction of the $Z_1$ boson in the Higgs-boson rest frame. The spin analyzers are given by the polar and azimuthal angles $(\theta, \phi)$ of each negatively charged lepton in its respective parent's rest frame.
The relevant off-diagonal entanglement-sensitive parameters of the SDM are then obtained from products of spherical harmonics $Y_L^M(\theta,\phi)$ from event-by-event real-valued estimators~\cite{Aguilar-Saavedra:2022wam},
\[
c_{2\,M_{1}\,2\,M_{2}} = 40\pi\;\Re \left(
Y_{2}^{M_1}(\theta_1,\phi_1)\times Y_{2}^{M_2}(\theta_2,\phi_2)
\right),
\]
where the integers $M_1=-M_2$ are either 1 or 2.
The values of the SDM coefficients $C_{2\,M_{1}\,2\,M_{2}} = \langle c_{2\,M_{1}\,2\,M_{2}}\rangle$ are obtained from the averages of these event-by-event estimators over events within the kinematic region defined by the dilepton and four-lepton mass requirements listed in the previous paragraph. During the calculation of the average values corrections are applied for the effects of background processes, detector acceptance, reconstruction effects, and higher‑order electroweak corrections, as described in End Matter~\ref{sec:app:SDM}.
In the same‑flavor decay channels ($4e$, $4\mu$), quantum interference arising from multiple indistinguishable assignments of identical leptons to the intermediate $Z$ bosons leads to modifications of the SDM elements at the level of roughly 10\%~\cite{Aguilar-Saavedra:2024jkj}. These effects are corrected for as described in End Matter~\ref{sec:app:SDM}.

At leading order in perturbation theory and under the assumption described above, the density matrix of the $ZZ^*$ system has a well-defined structure with only nine of the 81 elements being non-zero~\cite{Aguilar-Saavedra:2022wam}.
Of the six off-diagonal elements, the four that are products of $\ket{0,0}$ and $\ket{\pm,\mp}$ are equal to $\frac{1}{3}C_{2,1,2,-1}$, while the two that are products of $\ket{+,-}$ and $\ket{-,+}$ are equal to $\frac{1}{3}C_{2,2,2,-2}$.
The Peres-Horodecki conditions~\cite{Peres:1996dw,Horodecki:1997vt} for the $ZZ^*$ system then imply that the system is entangled if and only if at least one of the coefficients $C_{2,1,2,-1}$ or $C_{2,2,2,-2}$ is non-zero~\cite{Aguilar-Saavedra:2022wam}.
A non-zero value of either of them serves as a marker of entanglement in the $ZZ^*$ system.
The null hypothesis in which both coefficients vanish corresponds physically to the situation in which the $Z$ bosons are in a pure separable $\ket{00}$ state,
with a single remaining non-zero SDM element.

\begin{figure}
\centering
\begin{subfigure}[b]{0.48\textwidth}
\includegraphics[width=\linewidth]{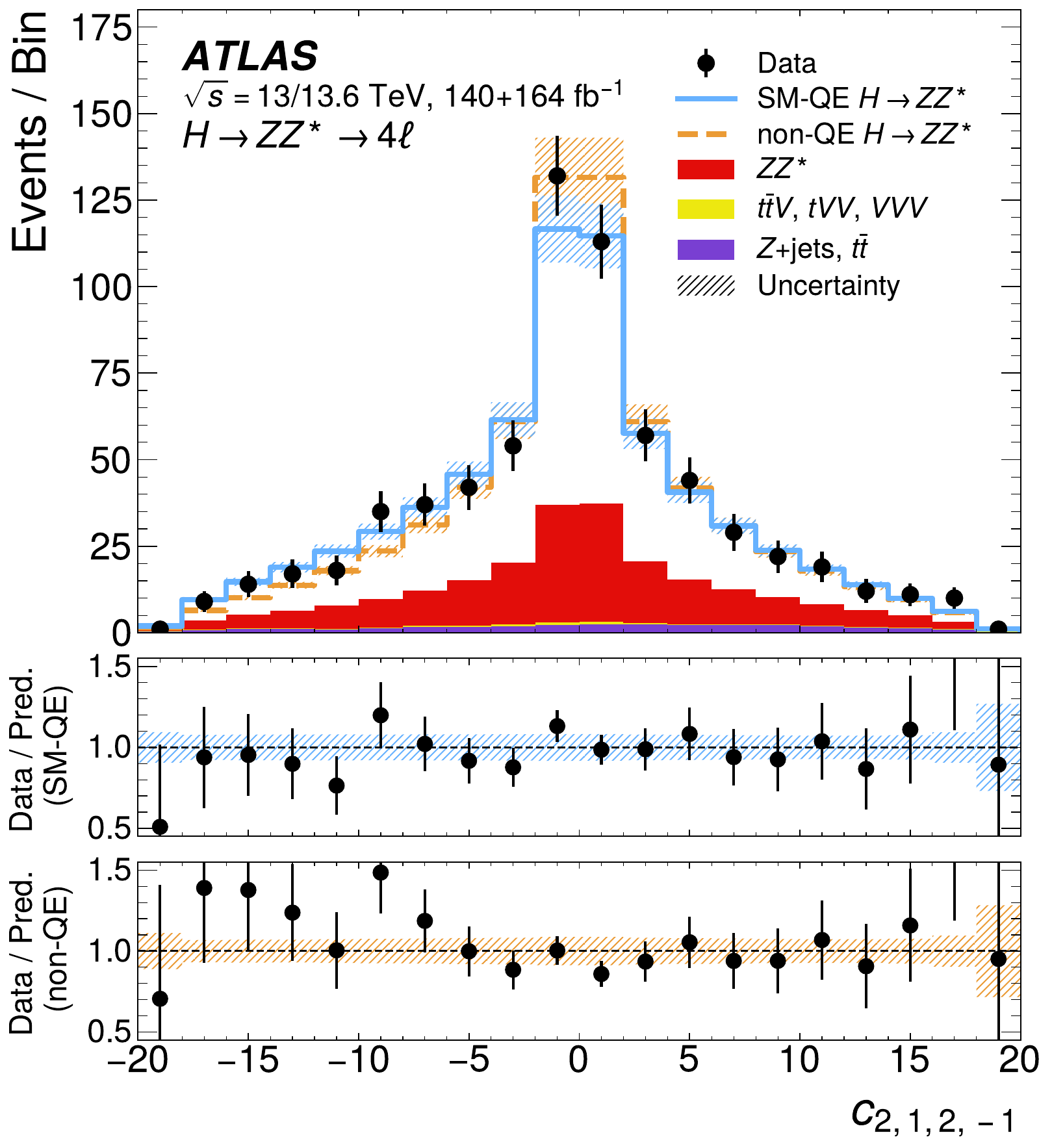}
\caption{$c_{2,1,2,-1}$}
\label{fig:c212-1,qe-nonqe}
\end{subfigure}
\hfill
\begin{subfigure}[b]{0.48\textwidth}
\includegraphics[width=\linewidth]{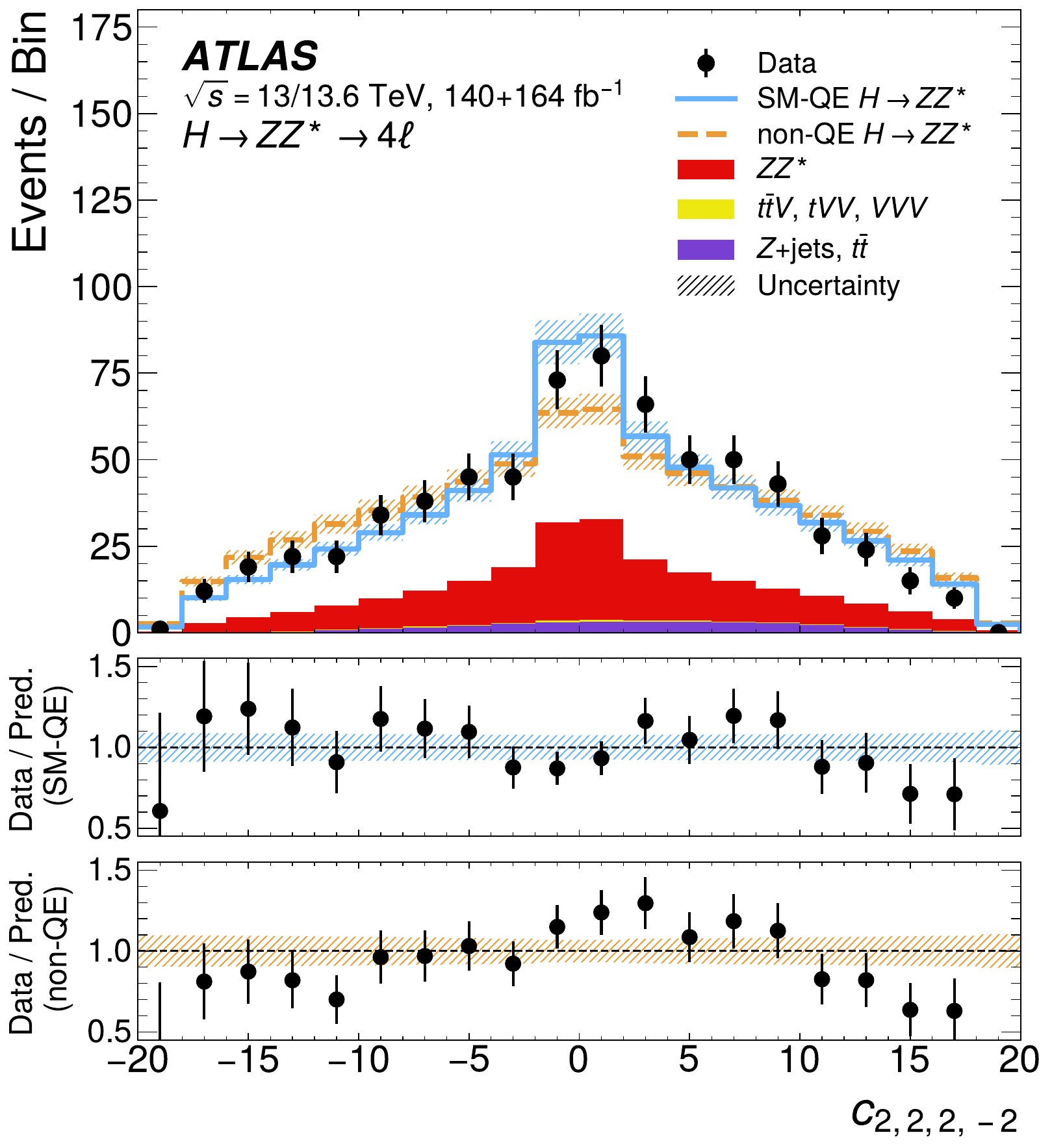}
\caption{$c_{2,2,2,-2}$}
\label{fig:c222-2,qe-nonqe}
\end{subfigure}
\caption{\label{fig:c222-2_c212-1,qe-nonqe}
{The observed distributions of events (full circles) overlaid on the expected (shaded) distributions of the estimator \subref{fig:c212-1,qe-nonqe}~$c_{2,1,2,-1}$ and \subref{fig:c222-2,qe-nonqe}~$c_{2,2,2,-2}$
for the entangled hypothesis (blue solid line) and the separable non-QE hypothesis (orange dashed line) and background.
}
}
\end{figure}

The sensitivity to entanglement from measurements of the coefficients $C_{2,1,2,-1}$ and $C_{2,2,2,-2}$ is limited, since they rely on estimating the mean of a broad underlying distribution using the mean of the data sample. To address this, an alternative approach is employed: the hypothesis of quantum entanglement (QE) as predicted by the Standard Model is tested against the null hypothesis of a separable system.
This null hypothesis corresponds to the $\ket{00}$ state, the only separable (``non-QE'') quantum state consistent with the stated assumptions -- transitions to the other possible separable states $\ket{+-}$ or $\ket{-+}$ would violate CP.
The full distribution of the $c_{2,2,2,-2}$ observable is used, which provides about 20\% higher expected sensitivity than using $c_{2,1,2,-1}$.

Based on the expected distributions of the $c_{2,2,2,-2}$ observable for each of the two hypotheses, large ensembles of pseudo-experiment datasets are produced,  with event yields in each analysis bin sampled according to Poisson statistics around the expected values.
All theoretical and experimental systematic uncertainties enter as nuisance parameters (NPs) with Gaussian constraints.
The test statistic for the hypothesis test
$\tilde{q} = -2 \ln \left(\mathcal{L}_\mathrm{\ket{00}} / \mathcal{L}_\mathrm{QE} \right)
$
is then computed for each pseudo-experiment by profiling the ratio of the two likelihoods
$\mathcal{L}_\mathrm{\ket{00}}$ and $\mathcal{L}_\mathrm{QE}$
over the NPs, with the overall normalization of the $H\rightarrow Z Z^*\rightarrow 4\ell$ process being treated as a free parameter.
The likelihood fit is performed simultaneously in two separate signal regions defined by requirements on the dilepton mass: $m_{Z_2}< 30\,\GeV$ and  $m_{Z_2}>30\,\GeV$. This additional event classification makes better statistical use of the predicted larger contribution of transverse polarisation components in the SM at higher $m_{Z_2}$.
Within the stated assumptions regarding the scalar nature of the Higgs boson and the SM chiral couplings in the $Z$ boson decays, the $m_{Z_2}$ spectrum for the purely longitudinal $\ket{00}$ configuration, the only separable state allowed, is expected to be shifted towards lower values compared to the SM entanglement configuration~\cite{Aguilar-Saavedra:2022mpg}, so the $m_{Z_2}$ distribution also indirectly probes the entanglement of the $ZZ^*$ system.
The split of the signal region leads to an improvement in sensitivity of about 15\% in the rejection of the purely longitudinal configuration.
The improvement arises primarily from differences in the expected angular distributions across the different $m_{Z_2}$ selections, while the $m_{Z_2}$ distribution itself contributes noticeably less.
The resulting $\tilde{q}$ distributions for the simulations of the QE and the $\ket{00}$ hypotheses are compared to determine the confidence level for rejecting the non-entangled hypothesis using the $\mathrm{CL}_{s}$ method~\cite{Read:2002hq}.

\begin{figure}
\centering
\begin{subfigure}[b]{0.48\textwidth}
\includegraphics[width=\linewidth]{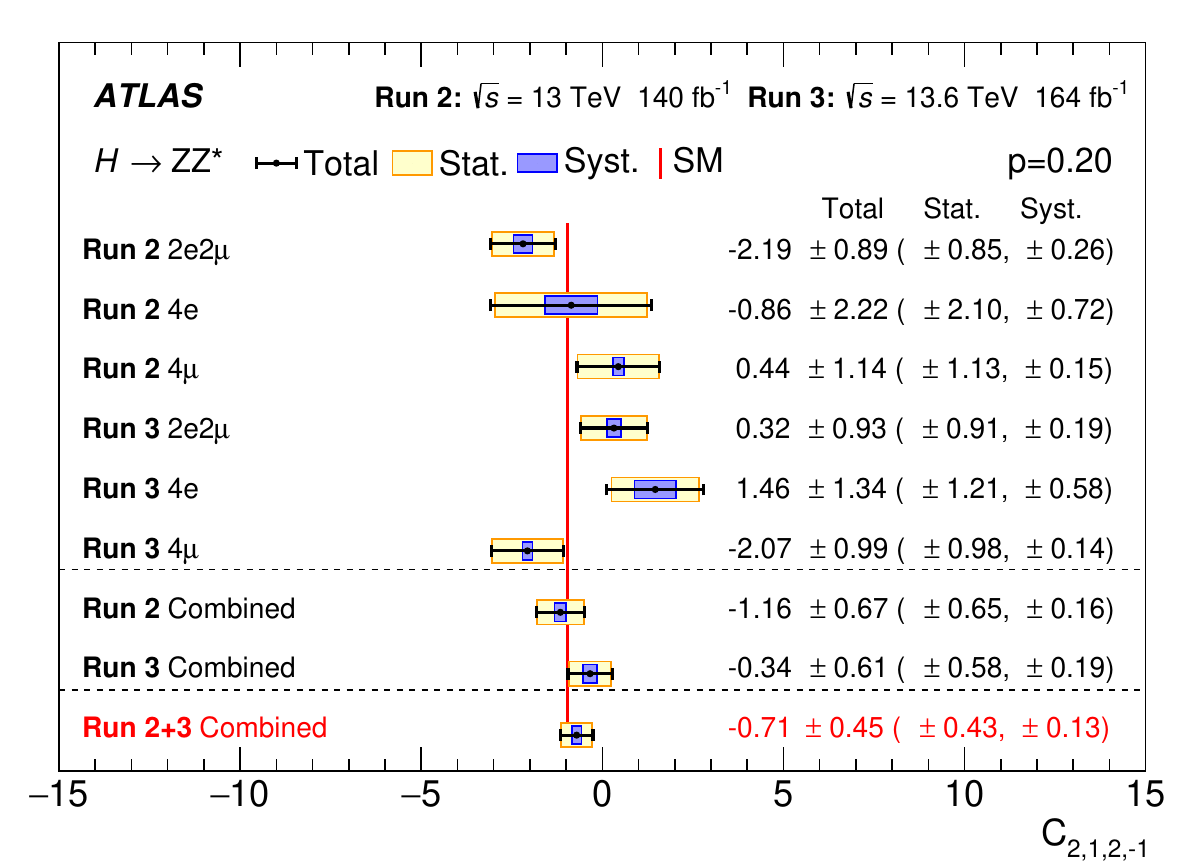}
\caption{$C_{2,1,2,-1}$}
\label{fig:C_combination:c212-1}
\end{subfigure}
\hfill
\begin{subfigure}[b]{0.48\textwidth}
\includegraphics[width=\linewidth]{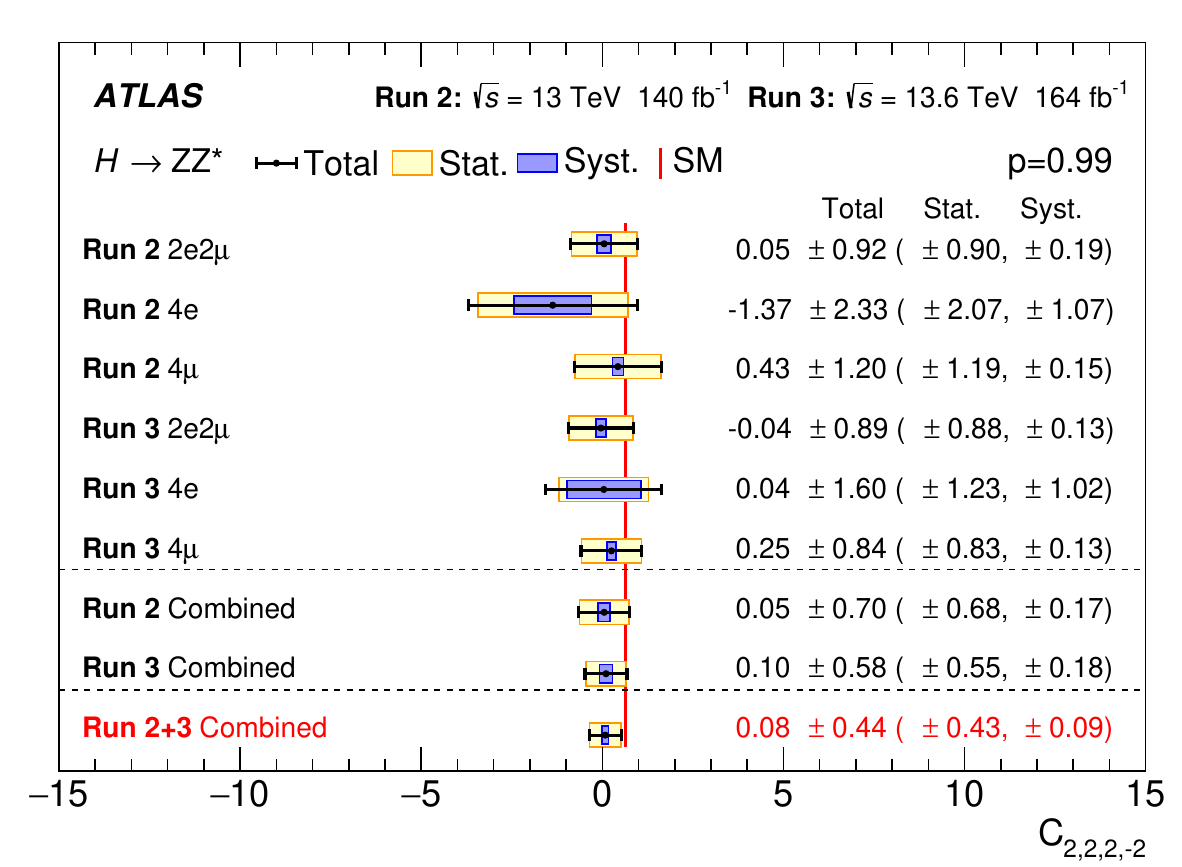}
\caption{$C_{2,2,2,-2}$}
\label{fig:C_combination:c222-2}
\end{subfigure}
\caption{
Observed values of \subref{fig:C_combination:c212-1} $C_{2,1,2,-1}$  and \subref{fig:C_combination:c222-2} $C_{2,2,2,-2}$ in the Run~2 and Run~3 datasets and their combination, together with the corresponding individual measurements. The $p$-values are calculated using the $\chi^2$ statistic for the six channels.
}
\label{fig:C_combination}
\end{figure}

The event-by-event distributions of the entanglement-sensitive observables $c_{2,2,2,-2}$ and $c_{2,1,2,-1}$ are shown in Figure~\ref{fig:c222-2_c212-1,qe-nonqe}, inclusively for the combination of both $Z_2$ signal regions.
The agreement between data and prediction for the relevant mass and angular distributions has been validated in previous measurements~\cite{HIGG-2018-29,ATLAS:2026hsl}.
The resulting parton-level SDM coefficients, %
together with their statistical and systematic uncertainties, are evaluated separately for the three decay channels ($4e$, $4\mu$, $2e2\mu$) and for the two data sets (Run~2, Run~3) and are then combined for the entanglement interpretation.
A summary of all six individual measurements and their combinations is shown in Figure~\ref{fig:C_combination}, separately for the
$C_{2,1,2,-1}$ and $C_{2,2,2,-2}$ coefficients.
The combinations are implemented using the Best Linear Unbiased Estimate technique~\cite{BLUE_LYONS1988110}.
The observed combined values of the coefficients,
shown in Table~\ref{tab:uncertainties} together with their uncertainties,
are found to be consistent between channels and consistent with the Standard Model predictions.
The systematic uncertainties
are all substantially smaller than the statistical uncertainties.

\begin{table*}
\caption{\label{tab:uncertainties}%
The observed and expected values of the entanglement-sensitive spin-density-matrix coefficients are shown, together with their dominant systematic and statistical uncertainties.
The `combined' columns show the results for $2e2\mu + 4e + 4\mu$.
The results are for the combined datasets from Run~2 and Run~3.
The observed and predicted values for the $4e + 4\mu$ channels are shown after the corrections $\Delta C^{\mathrm{interf}}$ for interference effects have been applied as described in End Matter~\ref{sec:app:SDM}. The SM expected values are calculated from the Monte Carlo simulation.
}
\renewcommand\arraystretch{1.2}
\centering\small
\begin{tabular}{|l|r|r|r|r|r|r|}
\hline
Coefficient
& \multicolumn{3}{c|}{$C_{2,1,2,-1}$}
& \multicolumn{3}{c|}{$C_{2,2,2,-2}$} \\
\hline
Channel & \multicolumn{1}{c|}{$2e2\mu$}
& \multicolumn{1}{c|}{$4e + 4\mu$}
& \multicolumn{1}{c|}{Combined}
& \multicolumn{1}{c|}{$2e2\mu$}
& \multicolumn{1}{c|}{$4e + 4\mu$ }
& \multicolumn{1}{c|}{Combined}\\
\hline
Observed
& $-0.98 \pm 0.64$
& $-0.44 \pm 0.63$
& $-0.71 \pm 0.45$
& $0.00 \pm 0.64$
& $0.15 \pm 0.61$
& $0.08 \pm 0.44$ \\
\hline
SM expected
& \multicolumn{3}{c|}{$-0.97$}
& \multicolumn{3}{c|}{$0.64$}\\
\hline
\multicolumn{7}{|l|}{Leading uncertainty components}\\
\hline
Fake-lepton bkg.
& $\pm 0.16$
& $\pm 0.16$
& $\pm 0.12$
& $\pm 0.11$
& $\pm 0.18$
& $\pm 0.12$ \\
Irreducible bkg.
& $\pm 0.05$
& $\pm 0.04$
& $\pm 0.04$
& $\pm 0.01$
& $\pm 0.01$
& $\pm 0.01$ \\
Interference
& ---
& $\pm 0.04$
& $\pm 0.02$
& ---
& $\pm 0.03$
& $\pm 0.02$ \\
EW NLO effects
& $\pm 0.02$
& $\pm 0.01$
& $\pm 0.01$
& $<$0.01
& $<$0.01
& $<$0.01 \\
Migration
& $\pm 0.01$
& $\pm 0.02$
& $\pm 0.01$
& $\pm 0.01$
& $\pm 0.02$
& $\pm 0.01$ \\
Statistical
& $\pm 0.62$
& $\pm 0.61$
& $\pm 0.43$
& $\pm 0.63$
& $\pm 0.58$
& $\pm 0.43$ \\
\hline
\end{tabular}
\end{table*}

\begin{figure}
\centering
\includegraphics[width=0.5\linewidth]{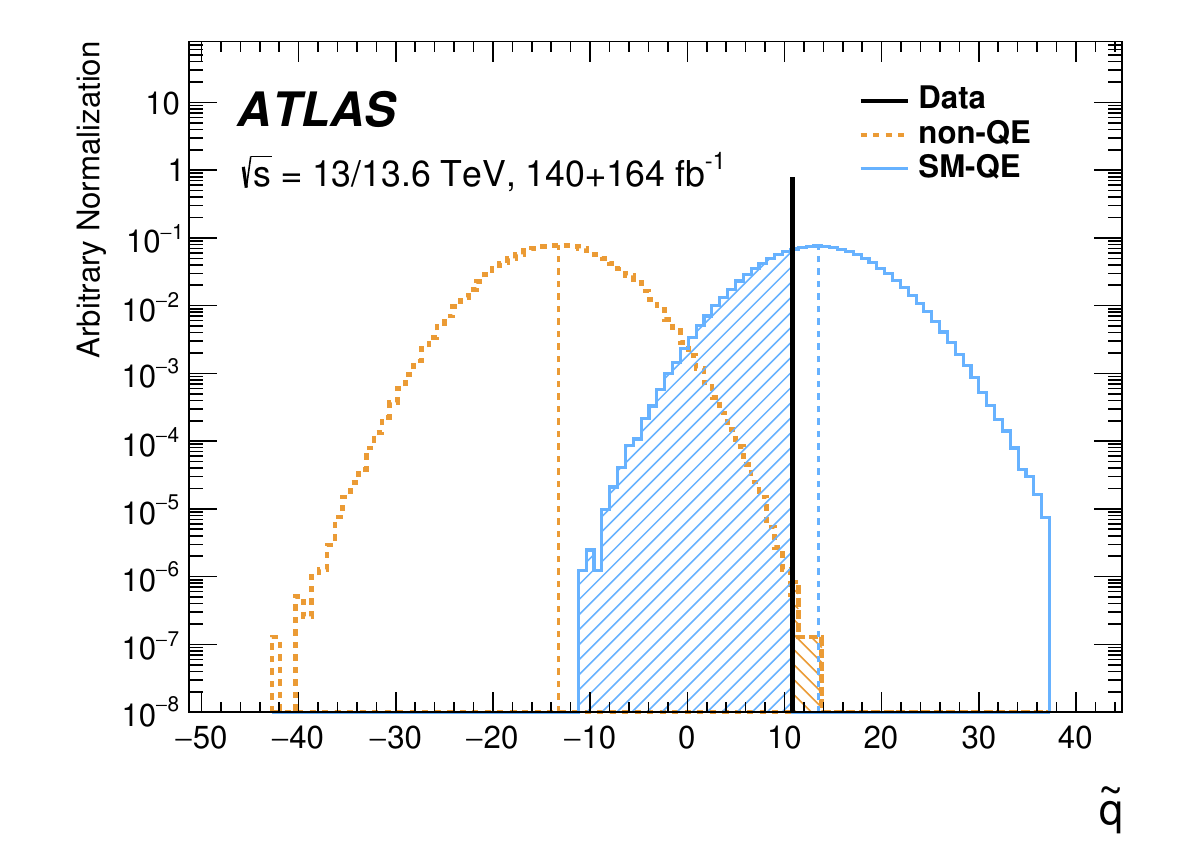}
\caption{\label{fig:hypo-qtilde}
Distribution of the test statistic $\tilde q$ for the entangled (blue solid line) and non-entangled (orange dashed line) hypotheses, as obtained from 10 million pseudo-experiments. The observed value is indicated by a vertical full line, and the expected values for each of the two hypotheses are indicated with vertical dashed lines.}
\end{figure}

The likelihood-ratio test has substantially higher sensitivity to the entangled SM hypothesis relative to the separable $\ket{00}$ hypothesis because it makes use of the full shape of the $c_{2,2,2,-2}$ distribution, rather than just its mean value.
The expected distributions of the test statistic $\tilde{q}$ for the two hypotheses are shown in Figure~\ref{fig:hypo-qtilde}, together with the observed value.
With all uncertainties taken into account, and based on 10 million pseudo-experiments per hypothesis,
the hypothesis of a non-entangled quantum state with purely longitudinally polarized $Z$ bosons is rejected with a significance of $4.7\sigma$ (expected $4.9\sigma$) in favor of the quantum-entanglement hypothesis predicted by the Standard Model.

In conclusion, these results constitute the first measurements of quantum entanglement
between two massive vector bosons at the electroweak scale.
The distinctive chiral structure of the SM weak force provides access to the $ZZ^*$ SDM in $H\rightarrow ZZ^*\rightarrow\ell^+\ell^-\ell^+\ell^-$ decays.
The results show how quantum coherence, previously observed in fermionic systems such as top-quark pairs,
extends to heavy spin-1 bosons in fundamental particle interactions.
With the larger datasets expected from the full Run~3 and the high-luminosity LHC,
future analyses will enable more precise tests
and sensitivity to potential deviations from the Standard Model.

\FloatBarrier
\section*{Acknowledgments}
%

%

%
%

%
%

We thank CERN for the very successful operation of the LHC and its injectors, as well as the support staff at
CERN and at our institutions worldwide without whom ATLAS could not be operated efficiently.

The crucial computing support from all WLCG partners is acknowledged gratefully, in particular from CERN, the ATLAS Tier-1 facilities at TRIUMF/SFU (Canada), NDGF (Denmark, Norway, Sweden), CC-IN2P3 (France), KIT/GridKA (Germany), INFN-CNAF (Italy), NL-T1 (Netherlands), PIC (Spain), RAL (UK) and BNL (USA), the Tier-2 facilities worldwide and large non-WLCG resource providers. Major contributors of computing resources are listed in Ref.~\cite{ATL-SOFT-PUB-2026-001}.

We gratefully acknowledge the support of ANPCyT, Argentina; YerPhI, Armenia; ARC, Australia; BMWFW and FWF, Austria; ANAS, Azerbaijan; CNPq and FAPESP, Brazil; NSERC, NRC and CFI, Canada; CERN; ANID, Chile; CAS, MOST and NSFC, China; Minciencias, Colombia; MEYS CR, Czech Republic; DNRF and DNSRC, Denmark; IN2P3-CNRS and CEA-DRF/IRFU, France; SRNSFG, Georgia; BMFTR, HGF and MPG, Germany; GSRI, Greece; RGC and Hong Kong SAR, China; ICHEP and Academy of Sciences and Humanities, Israel; INFN, Italy; MEXT and JSPS, Japan; CNRST, Morocco; NWO, Netherlands; RCN, Norway; MNiSW, Poland; FCT, Portugal; MNE/IFA, Romania; MSTDI, Serbia; MSSR, Slovakia; ARIS and MVZI, Slovenia; DSI/NRF, South Africa; MICIU/AEI, Spain; SRC and Wallenberg Foundation, Sweden; SERI, SNSF and Cantons of Bern and Geneva, Switzerland; NSTC, Taipei; TENMAK, T\"urkiye; STFC/UKRI, United Kingdom; DOE and NSF, United States of America.

Individual groups and members have received support from BCKDF, CANARIE, CRC and DRAC, Canada; CERN-CZ, FORTE and PRIMUS, Czech Republic; COST, ERC, ERDF, Horizon 2020 and Marie Sk{\l}odowska-Curie Actions, European Union; Investissements d'Avenir Labex, Investissements d'Avenir Idex and ANR, France; DFG and AvH Foundation, Germany; Herakleitos, Thales and Aristeia programmes co-financed by EU-ESF and the Greek NSRF, Greece; BSF-NSF and MINERVA, Israel; NCN and NAWA, Poland; La Caixa Banking Foundation, CERCA and AGAUR programs from Generalitat de Catalunya and PROMETEO and GenT Programmes Generalitat Valenciana, Spain; G\"{o}ran Gustafssons Stiftelse, Sweden; The Royal Society and Leverhulme Trust, United Kingdom; Eric and Wendy Schmidt Fund for Strategic Innovation, United States of America.

In addition, individual members wish to acknowledge support from Chile: Agencia Nacional de Investigaci\'on y Desarrollo (ANID FONDECYT reg. 1230987, FONDECYT 1230812, FONDECYT 1240864, Fondecyt 3240661, Fondecyt Regular 1240721); China: Chinese Ministry of Science and Technology (MOST-2023YFA1605700, MOST-2023YFA1609300), National Natural Science Foundation of China (NSFC 12275265, NSFC-W2543005); Czech Republic: Czech Science Foundation (GACR - 24-11373S), Ministry of Education Youth and Sports (ERC-CZ-LL2327, FORTE CZ.02.01.01/00/22\_008/0004632), PRIMUS Research Programme (PRIMUS/21/SCI/017); EU: H2020 European Research Council (ERC - 101002463); European Union: European Research Council (BARD No. 101116429, ERC - 101219398, ERC - 948254, ERC 101089007), European Regional Development Fund (HE COFUND GA No.101081355, ERDF), Marie Sklodowska-Curie Actions (GAP-101168829); France: Agence Nationale de la Recherche (ANR-21-CE31-0013, ANR-22-EDIR-0002, ANR-24-CE31-0504-01); Germany: Deutsche Forschungsgemeinschaft (DFG - 469666862); China: Research Grants Council (GRF); Italy: Istituto Nazionale di Fisica Nucleare (LHC-MIUR - 28003/2025), Ministero dell'Università e della Ricerca (NextGenEU 153D23001490006 M4C2.1.1, NextGenEU I53D23000820006 M4C2.1.1, NextGenEU I53D23001490006 M4C2.1.1, SOE2024\_0000023); Japan: Japan Society for the Promotion of Science (JSPS KAKENHI  JP25H0063, JSPS KAKENHI JP22H01227, JSPS KAKENHI JP22H04944, JSPS KAKENHI JP22KK0227, JSPS KAKENHI JP24K23939, JSPS KAKENHI JP24KK0251, JSPS KAKENHI JP25H00650, JSPS KAKENHI JP25H01291, JSPS KAKENHI JP25K01011, JSPS KAKENHI JP25K01023); Poland: Polish National Science Centre (NCN 2021/42/E/ST2/00350, NCN OPUS 2023/51/B/ST2/02507, NCN OPUS nr 2022/47/B/ST2/03059, NCN UMO-2019/34/E/ST2/00393, UMO-2022/47/O/ST2/00148, UMO-2023/49/B/ST2/04085, UMO-2023/51/B/ST2/00920, UMO-2024/53/N/ST2/00869); Spain: Agència de Gestió d'Ajuts Universitaris i de Recerca. (AGAUR - 2023 BP 00141), Ministry of Science and Innovation (RYC2019-028510-I, RYC2020-030254-I, RYC2021-031273-I, RYC2022-038164-I), Ministerio de Ciencia, Innovación y Universidades/Agencia Estatal de Investigaci\'on (EU NextGenerationEU (PRTR-C17.I1), PID2022-142604OB-C22); Sweden: Carl Trygger Foundation (Carl Trygger Foundation CTS 22:2312), Swedish Research Council (Swedish Research Council 2023-04654, VR 2021-03651, VR 2022-03845, VR 2022-04683, VR 2023-03403, VR 2024-05451, VR 2025-05940), Knut and Alice Wallenberg Foundation (KAW 2023.0366); Switzerland: Swiss National Science Foundation (SNSF - PCEFP2\_194658); United Kingdom: The Binks Trust, Royal Society (NIF-R1-231091); United States of America: U.S. Department of Energy (ECA DE-AC02-76SF00515), John Templeton Foundation (John Templeton Foundation 63206), Neubauer Family Foundation.

%
%


\clearpage
\appendix

\clearpage
\part*{End Matter}

\section{Simulation and background prediction}\label{sec:app:MC}
Monte Carlo (MC) simulated signal samples for the Standard Model (SM) Higgs boson are simulated using
\textsc{Powheg~Box~v2}~\cite{Alioli:2010xd} interfaced to \textsc{Pythia~8}~\cite{Bierlich:2022pfr} with the dedicated \texttt{Hto4l}~\cite{Boselli:2017pef} decay library providing full spin correlations in $H\to ZZ^*\to 4\ell$ decays.
The dominant gluon--gluon fusion (ggF) process is simulated at next-to-next-to-leading order (NNLO) in quantum chromodynamics (QCD)~\cite{Hamilton:2013fea,Alioli:2021wpn},
while vector-boson fusion (VBF)~\cite{Nason:2009ai} and associated production ($VH$) are simulated at next-to-leading order (NLO).
Minor production modes of Higgs bosons in association with top and bottom quarks, including $t\bar{t}H$, $tH$, and $b\bar{b}H$, are simulated at NLO using \textsc{Powheg~Box~v2 + Pythia~8}.
Final-state radiation is simulated within \texttt{Hto4l} and \textsc{Pythia~8}. Its impact on the reconstructed angular observables was found to be below 1\%.

The samples are normalized to state-of-the-art cross-section predictions for the center-of-mass energy of 13\,\TeV~\cite{deFlorian:2016spz} and these predictions are extrapolated to 13.6\,\TeV.
The ggF sample is normalized to a N$^3$LO QCD calculation with NLO electroweak (EW) corrections~\cite{Aglietti:2004nj,Actis:2008ug,Actis:2008ts,Anastasiou:2008tj,Pak:2009dg,Harlander:2009bw,Harlander:2009mq,Harlander:2009my,Actis:2008ug,Anastasiou:2015ema,Anastasiou:2016cez,Dulat:2018rbf,Bonetti:2018ukf}.
The VBF sample is normalized to an approximate NNLO QCD
cross-section with NLO EW corrections~\cite{Ciccolini:2007jr,Ciccolini:2007ec,Bolzoni:2010xr}.
The $VH$ samples are normalized to cross-sections calculated at NNLO in
QCD with NLO EW corrections~\cite{Ciccolini:2003jy,Brein:2003wg,Brein:2011vx,Altenkamp:2012sx,Denner:2014cla,Brein:2012ne,Harlander:2014wda,Harlander:2018yio}.
The \bbH\ contribution is normalized to a calculation that combines the complete NLO contributions that are present in the 4-flavor scheme calculation, including finite $b$-quark mass effects and top-loop induced top-to-bottom Yukawa-coupling interference contributions, with the resummation of
collinear logarithms of $m_b/m_H$ as present in the 5-flavor scheme calculation up to NNLO~\cite{Dittmaier:2003ej, Dawson:2003kb, Wiesemann:2014ioa, Harlander:0304035, Harlander:2011aa}.
No EW corrections are included for \bbH, \tH or $tWH$.

Irreducible backgrounds from non-resonant $Z^{(*)}Z^{(*)}$ production are modeled using \textsc{Sherpa}~2.2.16~\cite{Bothmann:2019yzt,Gleisberg:2008fv,Cascioli:2011va}, with a matrix element calculated at NLO in the strong coupling $\alpha_s$ for final states with up to one additional parton, and at LO accuracy for up to three additional partons.
The matrix element (ME) calculations used \texttt{OpenLoops}~\cite{Cascioli:2011va} for the virtual QCD corrections, and
were merged with the \textsc{Sherpa} parton shower using the ME+PS@NLO prescription~\cite{Hoeche:2012yf}.
The loop-induced gluon-initiated $ZZ$ process ($gg\rightarrow ZZ$) was simulated with \POWHEGBOX[v2]+ \PYTHIA[8] at NLO precision.

Production of $t\bar{t}Z$ and tri-boson ($VVV$) production, where $V$ represents $W^\pm$ or $Z$, are both modeled with \textsc{Sherpa}~2.2.12, requiring at least four charged leptons in the final state.
Backgrounds with non-prompt leptons arise primarily from $Z$ boson production with associated jets ($Z+$jets),
production of $WZ$ and $t\bar{t}$. The production of $Z+$jets was simulated using \textsc{Sherpa}~2.2.14 with up to two partons at NLO and five partons at LO in $\alpha_s$.
The $WZ$ process was simulated using \textsc{Sherpa}~2.2.16, with a similar configuration as used for the $q\bar{q}\rightarrow Z^{(*)}Z^{(*)}$ process.
Events of $t\bar{t}$ production were simulated with \POWHEGBOX[v2]+\PYTHIA[8.3] at NLO in $\alpha_s$.

For the test of the quantum-entanglement hypothesis, additional dedicated samples are produced using \textsc{MadGraph5\_aMC@NLO}~\cite{Alwall:2014hca} enforcing longitudinal polarization of both $Z^{(*)}$ bosons.
They are parton-showered with \textsc{Pythia~8} and reweighted
to match the
production kinematic spectra of the \textsc{Powheg}-based inclusive SM Higgs-boson samples.

The reducible background arises when events with fewer than four prompt leptons satisfy the signal region event selection criteria because jets or photons are misidentified as leptons or due to non-prompt leptons from heavy-flavor hadron decays. Since such backgrounds may not be well-modeled in simulation, this contribution is estimated with a data-driven fake-factor (FF) method~\cite{EGAM-2019-01} rather than from direct MC simulation.
The FF is measured in two dedicated control regions enriched respectively in $Z$+jets (CR $Z$) and in top-pair production (CR top).
Systematic uncertainties include the variation of the control region for the FF measurement (CR Z or CR top), the statistical uncertainty in the measurement CRs, and the statistical uncertainty in the application regions.
The total uncertainty in the reducible background estimate in the signal region is 39\% for the Run~2 data and 46\% for the Run~3 data.

\section{Calibration corrections and systematic uncertainties}
\label{sec:app:SDM}

The procedure for evaluating the SDM elements is as follows. First, the mean value $C^\mathrm{data, reco}_\mathrm{ch}$ of the entanglement-sensitive angular distribution $c^\mathrm{data, reco}_\mathrm{ch}$ is calculated using all events that satisfy the event selection criteria. The index `$\mathrm{ch}$' denotes the individual decay channel, $4e$, $4\mu$, or $2e2\mu$.
Next, the reconstructed mean value $C^\mathrm{data, reco}_\mathrm{ch}$ is corrected for background effects, using their predicted yields and mean values of the coefficient, resulting in a reconstructed value $C^{\mathrm{H4\ell,reco}}_{\mathrm{ch}}$ for the Higgs-boson signal alone.
The statistical and systematic uncertainty in all background contributions are propagated.
The reconstructed coefficient $C^\mathrm{H4\ell, reco}_\mathrm{ch}$ is calibrated to the `parton' (i.e. leptons before QED radiation) level value using a set of linear calibrations, one for each channel for each LHC run period (i.e Run~2 and Run~3), obtained from $c^\mathrm{H4\ell, MC}_\mathrm{ch}$ weighted simulations of Standard Model Higgs-boson signal samples. The systematic uncertainty of the calibration procedure is evaluated and taken into account.
For same-flavor decay channels,  the correction $\Delta C^{\mathrm{interf}} =  C^\mathrm{MC}_{2e2\mu} - C^\mathrm{MC}_{4e/4\mu}$ is applied to account for interference effects, where the corrections are found from simulation to be
$\Delta C_{2,1,2,-1}^{\mathrm{interf}} = -0.07$
and
$\Delta C_{2,2,2,-2}^{\mathrm{interf}} = 0.06$.
A conservative symmetrical systematic uncertainty equal to $50\%$ of this correction is assigned to the same-flavor channels (4e and 4$\mu$).
Finally, an additional multiplicative systematic uncertainty is introduced to account for effects from electroweak corrections at NLO. This uncertainty is taken as the difference between LO and NLO EW predictions~\cite{Aguilar-Saavedra:2025byk}, corresponding to multiplicative uncertainties of $1.1\%$ for $C_{2,1,2,-1}$ and $1.4\%$ for $C_{2,2,2,-2}$.

Experimental systematic uncertainties are accounted for including lepton calibration and acceptance, luminosity, as well as the effect of pile-up~\cite{HIGG-2020-07}. Uncertainties in the yield of fake- and non-prompt leptons are estimated using dedicated control regions.
For the combined measurement of each SDM coefficient, the statistical uncertainties, including those on the fake-lepton estimate, and the systematic uncertainty from unfolding are each treated as uncorrelated between flavor channels and between LHC run periods.
The uncertainties from biases from the interference (where present) and those from NLO EW correction are treated as fully correlated between flavor channels and run periods. The systematic uncertainties resulting from the background subtraction and from the fake-lepton estimate closure are treated as fully correlated between flavor channels and uncorrelated between run periods.

For the hypothesis test, the overall impact from all sources of systematic uncertainties is indicated by the uncertainty band in Figure~\ref{fig:c222-2_c212-1,qe-nonqe}. The dominant theoretical uncertainty originates from the theoretical contribution from higher-order terms in QCD, and corresponds to approximately 10\% of the predicted yield. Sub-leading systematic uncertainties include those from the MC parton shower modeling and reducible background (fake-lepton) related uncertainties. Experimental uncertainties are found to be subdominant and are found not to materially change the shape of the distributions. They are therefore represented by a conservative overall 3\% uncertainty.

\clearpage

\newpage
\printbibliography

\clearpage
 
\begin{flushleft}
\hypersetup{urlcolor=black}
{\Large The ATLAS Collaboration}

\bigskip

\AtlasOrcid[0000-0002-6665-4934]{G.~Aad}$^\textrm{\scriptsize 102}$,
\AtlasOrcid[0000-0001-7616-1554]{E.~Aakvaag}$^\textrm{\scriptsize 17}$,
\AtlasOrcid[0000-0002-5888-2734]{B.~Abbott}$^\textrm{\scriptsize 121}$,
\AtlasOrcid[0000-0002-0287-5869]{S.~Abdelhameed}$^\textrm{\scriptsize 83b}$,
\AtlasOrcid[0000-0002-1002-1652]{K.~Abeling}$^\textrm{\scriptsize 54}$,
\AtlasOrcid[0000-0001-5763-2760]{N.J.~Abicht}$^\textrm{\scriptsize 48}$,
\AtlasOrcid[0000-0002-8496-9294]{S.H.~Abidi}$^\textrm{\scriptsize 30}$,
\AtlasOrcid[0009-0003-6578-220X]{M.~Aboelela}$^\textrm{\scriptsize 44}$,
\AtlasOrcid[0000-0002-9987-2292]{A.~Aboulhorma}$^\textrm{\scriptsize 36e}$,
\AtlasOrcid[0000-0001-5329-6640]{H.~Abramowicz}$^\textrm{\scriptsize 154}$
% ...
\bigskip
\\

% ...
$^{17}$Department for Physics and Technology, University of Bergen, Bergen; Norway.\\
% ...
$^{30}$Physics Department, Brookhaven National Laboratory, Upton NY; United States of America.\\
% ...
$^{36}$$^{(a)}$Facult\'e des Sciences Ain Chock, Universit\'e Hassan II de Casablanca;$^{(b)}$Facult\'{e} des Sciences, Universit\'{e} Ibn-Tofail, K\'{e}nitra;$^{(c)}$Facult\'e des Sciences Semlalia, Universit\'e Cadi Ayyad, LPHEA-Marrakech;$^{(d)}$LPMR, Facult\'e des Sciences, Universit\'e Mohamed Premier, Oujda;$^{(e)}$Facult\'e des sciences, Universit\'e Mohammed V, Rabat;$^{(f)}$Institute of Applied Physics, Mohammed VI Polytechnic University, Ben Guerir; Morocco.\\
% ...
$^{44}$Physics Department, Southern Methodist University, Dallas TX; United States of America.\\
% ...
$^{48}$Fakult\"{a}t Physik, Technische Universit{\"a}t Dortmund, Dortmund; Germany.\\
% ...
$^{54}$II. Physikalisches Institut, Georg-August-Universit\"{a}t G\"ottingen, G\"ottingen; Germany.\\
% ...
$^{83}$$^{(a)}$Khalifa University of Science and Technology, Abu Dhabi;$^{(b)}$New York University Abu Dhabi, Abu Dhabi;$^{(c)}$United Arab Emirates University, Al Ain;$^{(d)}$University of Sharjah, Sharjah; United Arab Emirates.\\
% ...
$^{102}$CPPM, Aix-Marseille Universit\'e, CNRS/IN2P3, Marseille; France.\\
% ...
$^{121}$Homer L. Dodge Department of Physics and Astronomy, University of Oklahoma, Norman OK; United States of America.\\
% ...
$^{154}$Raymond and Beverly Sackler School of Physics and Astronomy, Tel Aviv University, Tel Aviv; Israel.\\
% ...
$^{a}$ Also at Affiliated with an institute formerly covered by a cooperation agreement with CERN.\\
$^{b}$ Also at An-Najah National University, Nablus; Palestine.\\
$^{c}$ Also at Borough of Manhattan Community College, City University of New York, New York NY; United States of America.\\
$^{d}$ Also at Center for Interdisciplinary Research and Innovation (CIRI-AUTH), Thessaloniki; Greece.\\
$^{e}$ Also at Centre of Physics of the Universities of Minho and Porto (CF-UM-UP); Portugal.\\
$^{f}$ Also at CERN, Geneva; Switzerland.\\
% ...

\end{flushleft}


%

%

\end{refsection}

\end{document}